\documentclass{article}
\usepackage{graphicx} % Required for inserting images
\usepackage{natbib}
\usepackage{xcolor}
\usepackage{url}
\usepackage{alphabeta}
\usepackage{amssymb}
\usepackage{amsmath,bm,calc,float}
\usepackage{hyperref}
\usepackage{subfigure}
\usepackage{graphicx}
\usepackage{comment}
\hypersetup{
    colorlinks=true,
    linkcolor=blue,
    citecolor=blue
    } %chang
\usepackage[hmarginratio=1:1]{geometry}
\usepackage{tablefootnote}

\title{Beyond Equidistant Assumptions: An Autoregressive Ordered Stereotype Model for Ordinal Time Series \\  [5ex]} 
\author{
Anna Nalpantidi$^{1}$,
Dimitris Karlis$^{1}$\footnote{Corresponding author: Address: 76 Patission str, 10434 Athens, Greece, karlis@aueb.gr, Tel: +30 210 8203920},
Daniel Fernández$^{2}$\\[1ex]
$^{1}$Department of Statistics, \\Athens University of Economics and Business\\ [1ex]
\\
$^{2}$Institute for Research and Innovation in Health (IRIS),\\ Universitat Politècnica de Catalunya - \\
BarcelonaTech (UPC), 08028 Barcelona, Spain \\ [1ex]}
\date{June 2026}

\begin{document}

\maketitle

%\newpage

%\end{document}

%\begin{center}
 %   \LARGE{Beyond Equidistant Assumptions: An Autoregressive Ordered Stereotype Model for Ordinal Time Series}
%\end{center}

%\vspace{2cm}

\begin{abstract}
We propose an extension of the ordered stereotype model (OSM) for ordinal time series data, referred to as the Autoregressive OSM (AR-OSM). The model captures serial dependence by incorporating lagged values of the response as covariates in the systematic component. In contrast to existing regression models for ordinal time series, the AR-OSM does not assume equidistant categories, but instead allows the data to determine their relative spacing. This property makes the model particularly suitable for applications where the equidistance assumption is unrealistic. Such a case is illustrated through the analysis of infant sleep state data. Additionally, a comprehensive simulation study is conducted to assess the performance of the model under varying sample sizes and to investigate how parameter values influence the induced serial dependence structure. 
\end{abstract}


\section{Introduction}

 Consider the sleep data in   \cite{fokianos2003regression}.
The data refer to the sleep states of a newborn infant. More specifically, EEG sleep state scores of a newborn infant are classified into four categories: (1) quiet sleep, (2) indeterminate sleep, (3) active and (4): awake. These states have a natural ordering from awake to active sleep, which can be expressed as $(4) < (1) < (2) < (3)$.
For this particular dataset, we have also available some covariate information like the temperature and heart rate every $30$ seconds for a newborn infant. The data constitute a time series with ordinal data. The different states are not necessarily equi-distant and it would be nice to be able to model with an ordinal time series model that can also infer the distance between the states. This is the purpose of the paper to extend the ordered stereotype model (OSM) so as to be able to model ordinal time series data.

Ordinal time series is a sequence of observations $Y_t,~ t=1,\ldots,T$ that takes values in a finite number of $q$ ordinal categories/states $\mathcal{S}=\{1,\ldots,q\}$. The key characteristic is that the states possess a natural ordering, that is, $1\le \cdots \le q$.
Ordinal time series arise in several fields such as environmental studies (\cite{gottlein1992ordinal}, \cite{liu2022modeling2}, \cite{liu2022modeling1}, \cite{jahn2024nonlinear}), sports \citep{fokianos2003regression}, healthcare \citep{fokianos2003regression}, and economics \citep{weiss2019distance}. For example, \cite{liu2022modeling2} investigated air quality across three major Chinese cities with the goal of improving pollution management. The daily air quality index for each city was recorded using a six-level ordinal scale: (1) excellent, (2) good, (3) slightly polluted, (4) moderately polluted, (5) heavily polluted, and (6) severely polluted, reflecting a clear natural ordering in the ordinal response.

%\textcolor{red}{In a different context, \cite{fokianos2003regression} analyzed sleep states of a newborn infant. More specifically, EEG sleep state scores of a newborn infant are classified into four categories: (1) quiet sleep, (2) indeterminate sleep, (3) active and (4): awake. These states have a natural ordering from awake to active sleep, which can be expressed as $(4) < (1) < (2) < (3)$. This dataset is also used to illustrate our proposed methodology. It includes sleep state score, temperature and heart rate every $30$ seconds for a newborn infant. Since the observations are clearly ordinal we aim at using our proposed model to fit such data and demonstrate the advantages it offers. }

The wide range of applications in which ordinal time series arise has motivated extensive research on their modeling, including Markov models \citep{raftery1985model,buhlmann1999variable}, regression models \citep{gottlein1992ordinal, pruscha1993categorical, fokianos2003regression}, latent continuous models \citep{varin2006pairwise}, 
discrete versions of ARMA models \citep{jacobs1978discrete, jacobs1978discrete2,jacobs1978discrete3, weiss2025weighted},
GARCH-type models \citep{weiss2023ordinal} or methods that treat ordinal variables as rank-count variables \citep{weiss2019distance,liu2022modeling2}. 

Among these approaches, ordinal regression models are considered an appealing choice, as they extend the familiar generalized linear models for ordinal responses in the context of time series. For instance, in the work of \cite{fokianos2003regression} popular models for ordinal time series like cumulative logistic model/proportional odds model and adjacent categories logit model are presented.  
Although the proportional odds model \citep{mccullagh1980regression} is very popular for analyzing ordinal data, it relies on the strong proportional assumption that explanatory variables have the same effect across all outcome categories. This proportionality assumption is often unrealistic in many real-world applications. More flexible alternatives, such as the partial proportional odds model \citep{peterson1990partial}, relax this assumption but at the cost of substantially increasing the number of parameters. 

The OSM introduced by \cite{anderson1984regression}, provides a flexible framework for modeling ordinal responses. It respects the ordinal nature of the data while allowing covariate effects to vary across categories in a structured manner, offering a compromise between the restrictive proportional odds model and more parameter-rich alternatives such as the partial proportional odds model.
This model can be interpreted as a constrained version of the baseline-category logit model, also known as the multinomial logistic regression model. While the latter treats the response as nominal, the OSM incorporates the ordinal structure through a set of monotone ``score'' parameters. This constraint yields a more parsimonious model than the fully general multinomial specification, while retaining sufficient flexibility to capture category-specific effects \citep{agresti2010analysis}.

A key feature of the OSM is that it allows the spacing between categories to be determined by the data, rather than assuming equidistant states as in many standard ordinal models. In contrast to the proportional odds model, which imposes a common effect of predictors across thresholds, the OSM accommodates varying effects across categories through this data-driven representation.

Despite its advantages, the model has not been as widely used as other ordinal regression models, due to the difficulty of its estimation that arises from the constraints of the ``score'' parameters. Nevertheless, parameter estimation of OSM can be addressed through standard maximum likelihood methods using the reparameterization described in \cite{fernandez2016mixture} to impose the monotone non decreasing constraint \eqref{Eq: constraint}. Works related to OSM can be found in \cite{fernandez2016mixture,fernandez2016goodness,fernandez2019method,fernandez2021archetypal,fernandez2020model}. Recently, the R package \texttt{clustord}~\cite{clustordpackage,
fernandezarnoldpledger2019} has been released, which provides an implementation of the OSM through the function \texttt{osm}, thereby facilitating its practical application.

While several models have been proposed for ordinal time series, many rely on restrictive assumptions such as proportional effects across categories or equidistant spacing between states. These assumptions may be unrealistic in practice, particularly in applications where the interpretation of distances between categories is meaningful. This motivates the need for more flexible modeling approaches that can account for both the ordinal structure and temporal dependence of the data. 

The main contribution of this work is to extend the ordered stereotype model to the setting of ordinal time series. We propose an Autoregressive OSM in which lagged values of the response are incorporated as covariates in the systematic component, allowing the model to capture serial dependence. In contrast to existing approaches for ordinal time series, the proposed model does not rely on the assumption of equidistant categories, but instead allows their spacing to be determined by the data. This provides a flexible and parsimonious framework for modeling ordinal time series with potentially unequal distances between states. \\

The paper is organized as follows. In \hyperref[Sec2]{Section 2}, we introduce the Autoregressive OSM of order $p$. Details about estimation and order selection are also included. In \hyperref[Sec3]{Section 3} the serial dependence, based on Ordinal Cohen's $\kappa$ is presented under different models. A simulation study to examine model’s performance under varying sample sizes and how different values of parameters affect the serial structure is presented in \hyperref[Sec4]{Section 4}. In \hyperref[Sec5]{Section 5} a real-world application involving infant sleep state data is studied. Finally, \hyperref[Sec6]{Section 6} summarizes the main findings and discusses further possible extensions.

\section{A new model for ordinal time series: The Autoregressive OSM (AR-OSM)}
\label{Sec2}
In this section, we present OSM and introduce the Autoregressive OSM. The AR-OSM is an extension of OSM appropriate for ordinal time series. It captures serial dependence by considering lagged values of itself in the systematic component of the regression. Then, response probabilities of each category are defined conditional to the previous states of the process. The order of the model, the lagged values to be considered, is specified through ordinal Cohen's $\kappa$ \citep{weiss2019distance} and its partial version.

\subsection{The OSM: Formulation}
\vspace{0.2cm}
\hrule \vspace{0.2cm}
\noindent
{\bf Definition 1: Ordered Stereotype Model (OSM)} 
Let $Y_i$ be an ordinal outcome with $q$ ordered categories for $i$ individual where $i=1,\ldots,n$. In addition, let $\bm{x}_i$ be a vector of predictors, either continuous or categorical, for the $i$ observation. Based on OSM, for the probability that $Y_i$ takes the category $k$ is characterized by the following log-odds: 
\begin{align*}
 \log \frac{P(Y_i = k \mid \bm{x}_i)}{P(Y_i = 1 \mid \bm{x}_i)} = \alpha_k + \phi_k(\bm{\beta}^T \bm{x}_i), \qquad i=1,\ldots,n, \qquad k = 2,\ldots,q,   
\end{align*}
where the inclusion of the following monotone increasing constraint
\begin{align}
    0 = \phi_1 \leq \phi_2 \leq \cdots \leq \phi_q = 1.
    \label{Eq: constraint}
\end{align}
to ensure that $Y_i$ is ordinal. 
$\bm{\beta}$ is a common regression coefficient vector that represents the effects of $\bm{x}$ on the log odds of the response variable for the category $k$ relative to the baseline category that is considered to be the first state. The parameters $\{\alpha_2,\ldots,\alpha_q\}$ are category-specific intercepts, the cut-points and  $\{\phi_1,\ldots,\phi_q\}$ are the parameters which can be
interpreted as the ``scores'' for the categories of the response variable
$Y_i$. To ensure identifiability of the model, we pose $\alpha_1=\phi_0$ and $\phi_k=1$. 
Then, under OSM the response probability of a category $k, ~k=2,\ldots,q$ is given by:
\begin{align*}
   P(Y_i = k \mid \bm{x}_i) = \frac{
\exp(\alpha_k + \phi_k(\bm{\beta}^T \bm{x}_i))
}{\sum_{\ell=1}^q \exp(\alpha_\ell + \phi_\ell(\bm{\beta}^T \bm{x}_i)}.
\end{align*}
while first category has probability
\begin{align*}
P(Y_i = 1 \mid \bm{x}_i) = \frac{1}{\sum_{\ell=1}^q \exp(\alpha_\ell + \phi_\ell(\bm{\beta}^T \bm{x}_i))}. 
\end{align*}
\vspace{0.2cm}
\hrule \vspace{0.2cm}
Parameters $\phi_k$ reveal the relative spacing of categories along a latent continuum. The distance between successive $\phi$'s reveal the distance of the ordered categories which we do not assume to be the same. If the $\phi_k$ are evenly spaced, the model is equivalent to the proportional odds version of the adjacent-categories logit model (see Chapter 4 of \cite{agresti2010analysis}).
More details about the model can be seen in \cite{fernandez2019method} and references therein.  Additionally, \citet{greenland1994} showed that the stereotype model is appropriate when the progression of the response variable occurs
through various stages.

\subsection{The AR-OSM: Formulation}

\vspace{0.2cm}
\hrule \vspace{0.2cm}
\noindent
{\bf Definition 2: Autoregressive Ordered Stereotype Model of order $p$} 

Let $(Y_t)_{t\in \mathcal{Z}}$ be an ordinal time series with state space $\mathcal{S}=\{1,\ldots,q\}$. The autoregressive ordered stereotype model of order $p$ for the probability of success of a $k$ category, $k\in\mathcal{S}$ is given by: 
\begin{align*}
    \log{\left(\frac{P(Y_t=k|\mathcal{F}_{t-1})}{P(Y_t=1|\mathcal{F}_{t-1})}\right)}&=\alpha_k+\phi_k\beta_1Y_{t-1}+\ldots+\phi_k\beta_p Y_{t-p}, ~~k=2,\ldots,q,
\end{align*}
where $\mathcal{F}_{t-1}=\bm{\sigma}\{Y_{i},i\leq t-1\}$ is the history of the time series, $\{\alpha_{1},\ldots,\alpha_{q}\}$ the set of category-specific intercepts, $\bm{\beta}=(\beta_1,\ldots,\beta_p)^{'}$ the vector of coefficients of autoregressive terms and $\{\phi_{1},\ldots,\phi_{q}\}$ the set of "score" parameters. We consider the first category to be the baseline category. For the "score" parameters, we have the constraint:
\begin{align*}
    0=\phi_{1}\leq\phi_{2}\leq \ldots\leq \phi_{q-1}\leq \phi_{q}=1, 
\end{align*}
to assure that $Y_t$ is an ordinal process. Similar to OSM,
identifiability is ensured by posing $\alpha_{1}=\phi_{1}=0$ and $\phi_{q}=1$. Parameters  $\phi_k$ represent the relative spacing of the ordinal categories.
Under AR-OSM, the response probability for a category  $k, ~k=2,\ldots,q$, conditional to the history of the process is given by:
\begin{align}
    P(Y_t=k|\mathcal{F}_{t-1})=\frac{\exp{(\alpha_k+\phi_k\beta_1Y_{t-1}+\ldots+\phi_k\beta_p Y_{t-p})}}{\sum_{\ell=1}^{q}\exp{(\alpha_\ell+\phi_\ell\beta_1Y_{t-1}+\ldots+\phi_\ell\beta_p Y_{t-p})}},
    \label{pk}
\end{align}
while for the first category it simplifies to the expression
\begin{align}
    P(Y_t=1|\mathcal{F}_{t-1})=\frac{1}{\sum_{\ell=1}^{q}\exp{(\alpha_\ell+\phi_\ell\beta_1Y_{t-1}+\ldots+\phi_\ell\beta_p Y_{t-p})}}. 
    \label{p1}
\end{align}
\vspace{0.2cm}
\hrule \vspace{0.2cm}

\noindent
\textbf{Remark:} The AR-OSM($p$) model can easily be extended for also including covariates. Let $\bm{X}_t=(X_{1t},\ldots,X_{St})'$ denotes a $S-$variate vector of covariates at time $t$, then the model can be adapted to allow for covariate information  in the following way: 
\begin{align*}
    \log{\left(\frac{P(Y_t=k|\mathcal{F}_{t-1})}{P(Y_t=1|\mathcal{F}_{t-1})}\right)}&=\alpha_k+\phi_k\beta_1Y_{t-1}+\ldots+\phi_k\beta_p Y_{t-p}+\gamma_1X_{1t}+\ldots+\gamma_KX_{St}, ~~k=2,\ldots,q,
\end{align*}

\subsection{Maximum Likelihood Estimation}
Estimation of the AR-OSM($p$) is based on the maximization of the conditional log-likelihood 
\begin{align*}
\ell(\bm{\theta})&=\sum_{t=p+1}^{T}\log(P(Y_t=y_t|\mathcal{F}_{t-1}))=\sum_{t=p+1}^{T}\log(P(Y_t=y_t|Y_{t-1}=y_{t-1},\ldots,Y_{t-p}=y_{y-p})),
\end{align*}
where $\bm{\theta}=(\alpha_2,\ldots,\alpha_q,\phi_2,\ldots,\phi_{q-1},\beta_1,\ldots,\beta_p)^{'}$ the vector of parameters. During the estimation procedure we need to pose the constraint that ensures that scores $\phi_1,\ldots,\phi_q$ are increasing. However, this restriction is complex to be imposed during maximization. To overcome it, we consider the reparameterization presented in \cite{fernandez2016mixture}. More specifically, we define $\mbox{v}_k=\mbox{logit}(\phi_k)$, for $k=2,\ldots,q-1$, that ensures 
\begin{align*}
    -\infty \leq \mbox{v}_1 \leq \mbox{v}_3 \leq \mbox{v}_2 \ldots \leq \mbox{v}_{q-1} \leq \infty. 
\end{align*}
Then, we set 
\begin{align*}
    \mbox{v}_k=\mbox{v}_{k-1}+\exp{(u_k)}, ~ \mbox{for}~~ -\infty < u_k < \infty~~ ~k=3,\ldots,q-1.  
\end{align*}
In this way we replace the constrained vector of parameters $\{\phi_1=0, \phi_2, ,\ldots,\phi_{q-1},\phi_q=1\}$ by the unconstrained $\{\mbox{v}_2,u_3,\ldots,u_{q-1}\}\in \mathbb{R}^{q-1}$. Once we find the MLEs of $\mbox{v}_2,u_3,\ldots,u_{q-1}$, we can transform back to the original set of parameters by
\begin{align*}
  \phi_k= \left\{
   \begin{array}{ll}
      0, & k=1\\
      \frac{1}{1+\exp{(-\mbox{v}_2)}}, & k=2\\
      \mbox{expit}\left[\mbox{logit}(\phi
      _2) + \sum_{\ell=3}^{k}\exp{(u_\ell)} \right], & k=3,\ldots,q-1 \\
      1, & k=q~,\\
\end{array} 
\right. 
\end{align*}
where $\mbox{expit}(x)=(1+\exp{(-x)})^{-1}$ the inverse of the logit function. 

Standard errors can be obtained from the Hessian matrix using the Delta method in the usual way. 

\subsection{Order selection}
 An essential issue when working with autoregressive models is the order specification. In common case of real-valued time series, Auto-correlation function (ACF) and Partial auto-correlation functions (PACF) can aid in model selection. For an AR($p$) model ACF typically decays gradually toward zero, while PACF will show significant peaks up to lag $p$ and then drop off to near zero. The idea is to use similar measures adapted in the context of ordinal time series. More specifically, \cite{weiss2019distance} presented an ordinal version of Cohen's $\kappa$ to measure serial dependence. 

\vspace{0.2cm}
\hrule \vspace{0.2cm}
\noindent
{\bf Definition 3: Ordinal Cohen's $\kappa$}
Let $Y_t$ be the ordinal time series with state space $\mathcal{S}=\{s_0,\ldots,s_d\}$ where it holds that $s_0<s_1<\ldots<s_d$. Then, assuming block distance between states, $d(s_i,s_j)=|i-j|$, the ordinal Cohen's $\kappa$, denoted by $\kappa_{o}(h)$ is given by:
\begin{align*}
    \kappa_{o}(h)=\frac{\sum_{i=0}^{d-1}(f_{ii}^{(h)}-f_i^2)}{\sum_{i=0}^{d}f_i(1-f_i)}
\end{align*}
where $f_i=P(Y_t\leq s_i)$ and  $f_{ii}^{(h)}=P(Y_t\leq s_i,Y_{t-h}\leq s_i)$, the joint cdf of $Y_t$ and $Y_{t-h}$, for $i=0,\ldots,d-1$.
\vspace{0.2cm}
\hrule \vspace{0.2cm}

Based on this, we propose a partial version of the ordinal Cohen's $\kappa$ to specify the autoregressive terms we need in the model. The idea behind partial ordinal Cohen's $\kappa$ is to be based on Durbin–Levinson Algorithm, like in PACF: 
\begin{align*}
    φ_{kk}&=\frac{\rho(k)-\sum_{j=1}^{k-1}φ_{k-1,j}\rho(k-j)}{1-\sum_{j=1}^{k-1}φ_{k-1,j}\rho(j)} \nonumber\\
    φ_{kj}&=φ_{k-1,j}-φ_{kk}φ_{k-1,k-j}, ~~j=1,\ldots,k-1
\end{align*}
where $φ_{kk}$ the PACF at lag $k$ and $\rho(k)$ the ACF at lag $k$, but instead we replace the ACF $\rho(k)$ with the ordinal Cohen's $\kappa$ $\kappa_{o}(k)$. In the work of \cite{weiss2018introduction},  partial versions of Cohen's $\kappa$  and Cramer's $u$ for categorical time series have already been presented in the same way.

\section{Properties: Ordinal Cohen's $\kappa$ for AR-OSM}
\label{Sec3}

\subsection{AR-OSM(1)}

Let $\kappa_{o}(h)$ denotes the ordinal Cohen's $\kappa$ at lag $h$ for a Markov chain $Y_t$ with state space $\mathcal{S}=\{1,2,.\ldots,q\}$. $Y_t$ is assumed to follow an AR-OSM($1$). In the following calculations we consider that $Y_t$ is stationary with stationary distribution $\bm{\pi}=(\pi_1,\ldots,\pi_q)^{'}$. Under stationarity marginal probabilities come from the stationary distribution. In addition we assume $Y_t$ as time-homogeneous, so in the transition matrix $P$ the elements $P_{ij}=P(Y_t=j|Y_{t-1}=i)$ -as specified in \eqref{pk}, \eqref{p1}—are constant over time which means that they do not depend on time:
$$P(Y_t=j|Y_{t-1}=i)=P(Y_s=j|Y_{s-1}=i),~~ \forall t,s$$
Moreover, the $h$-steps transition probabilities are given by
$$P_{ij}^{(h)}=P(Y_t=j|Y_{t-h}=i)=(P^h)_{ij}.$$

 For $h=1$, serial dependence is given  by: 

    \begin{align*}
        \kappa_o(1)&=\frac{\sum\limits_{i=1}^{q-1}(f^{(1)}_{ii}-f^{2}_i)}{\sum\limits_{i=1}^{q}f_i(1-f_i)}\\
        &=\frac{\sum\limits_{i=1}^{q-1} \Big[\Big( \sum\limits_{j_1=1}^{i}\sum\limits_{j_2=1}^{i} P(Y_t=j_2,Y_{t-1}=j_1)\Big)- (\sum\limits_{j=1}^{i}P(Y_t=j))^2\Big]}{\sum\limits_{i=1}^{q-1} \Big[(\sum\limits_{j=1}^{i}P(Y_t=j))(1-\sum\limits_{j=1}^{i}P(Y_t=j))\Big]}\\
        &=\frac{\sum\limits_{i=1}^{q-1} \Big[\Big( \sum\limits_{j_1=1}^{i}\sum\limits_{j_2=1}^{i} P(Y_t=j_2|Y_{t-1}=j_1)P(Y_{t-1}=j_1)\Big)- (\sum\limits_{j=1}^{i}P(Y_t=j))^2\Big]}{\sum\limits_{i=1}^{q-1}\Big[(\sum\limits_{j=1}^{i}P(Y_t=j))(1-\sum\limits_{j=1}^{i}P(Y_t=j))\Big]}\\
        &=\frac{\sum\limits_{i=1}^{q-1} \Big[\Big( \sum\limits_{j_1=1}^{i}\sum\limits_{j_2=1}^{i} P_{j_1j_2}\pi_{j_1}\Big)- (\sum\limits_{j=1}^{i}\pi_j)^2\Big]}{\sum\limits_{i=1}^{q-1} \Big[(\sum\limits_{j=1}^{i}\pi_j)(1-\sum\limits_{j=1}^{i}\pi_j)\Big]}
    \end{align*}
  while for the general case of $h>1$, the expression is
  \begin{footnotesize}
\begin{align*}
    &\kappa_{o}(h)=\frac{\sum\limits_{i=0}^{d-1}(f_{ii}^{(h)}-f_i^2)}{\sum\limits_{i=0}^{d}f_i(1-f_i)}\\
    &=\frac{\sum\limits_{i=1}^{q-1} \Big[\Big( \sum\limits_{j_1=1}^{i}\sum\limits_{j_2=1}^{i} P(Y_t=j_2,Y_{t-h}=j_1)\Big)- (\sum\limits_{j=1}^{i}P(Y_t=j))^2\Big]}{\sum\limits_{i=1}^{q-1} \Big[(\sum\limits_{j=1}^{i}P(Y_t=j))(1-\sum\limits_{j=1}^{i}P(Y_t=j))\Big]}\\
    &=\frac{\sum\limits_{i=1}^{q-1} \Big[\Big( \sum\limits_{j_1=1}^{i}\sum\limits_{j_2=1}^{i} P(Y_t=j_2|Y_{t-h}=j_1)P(Y_{t-h}=j_1)\Big)- (\sum\limits_{j=1}^{i}P(Y_t=j))^2\Big]}{\sum\limits_{i=1}^{q-1} \Big[(\sum\limits_{j=1}^{i}P(Y_t=j))(1-\sum\limits_{j=1}^{i}P(Y_t=j))\Big]}\\
    &=\frac{\sum\limits_{i=1}^{q-1} \Big[\Big( \sum\limits_{j_1,j_2=1}^{i} %\sum\limits_{j_2=1}^{i} 
    \Big(\sum\limits_{{\bm \ell} \in {\cal S}} P(Y_t=j_2|Y_{t-1}=\ell_1)P(Y_{t-1}=\ell_1|Y_{t-2}=\ell_2)\ldots P(Y_{t-h+1}=\ell_{h-1}|Y_{t-h}=j_2)\Big)P(Y_{t-h}=j_2)\Big)- (\sum\limits_{j=1}^{i}P(Y_t=j))^2\Big]}{\sum\limits_{i=1}^{q-1} \Big[(\sum\limits_{j=1}^{i}P(Y_t=j))(1-\sum\limits_{j=1}^{i}P(Y_t=j))\Big]}\\
   &=\frac{\sum\limits_{i=1}^{q-1} \Big[\Big( \sum\limits_{j_1,j_2=1}^{i}
   %\sum\limits_{j_2=1}^{i}
   P^{(h)}_{j_1j_2}\pi_j\Big)- (\sum\limits_{j=1}^{i}\pi_j)^2\Big]}{\sum\limits_{i=1}^{q-1} \Big[(\sum\limits_{j=1}^{i}\pi_j)(1-\sum\limits_{j=1}^{i}\pi_j)\Big]}
\end{align*}
\end{footnotesize}
In the above ${\bm \ell} = (\ell_1,\ldots,\ell_{h-1})$ with all $\ell_j \in \cal S$ so $\sum\limits_{\bm \ell}$ implies a multiple summation.

%    &=\frac{\sum\limits_{i=1}^{q-1} \Big[\Big( \sum\limits_{j_1=1}^{i}\sum\limits_{j_2=1}^{i} \Big(\sum\limits_{\ell_1 \in \mathcal{S}} \ldots \sum\limits_{\ell_{h-1} \in \mathcal{S}}P(Y_t=j_2|Y_{t-1}=\ell_1)P(Y_{t-1}=\ell_1|Y_{t-2}=\ell_2)\ldots P(Y_{t-h+1}=\ell_{h-1}|Y_{t-h}=j_2)\Big)P(Y_{t-h}=j_2)\Big)- (\sum\limits_{j=1}^{i}P(Y_t=j))^2\Big]}{\sum\limits_{i=1}^{q-1} \Big[(\sum\limits_{j=1}^{i}P(Y_t=j))(1-\sum\limits_{j=1}^{i}P(Y_t=j))\Big]}\\

\subsection{AR-OSM(2)}

Let $\kappa_{o}(h)$ denotes the ordinal Cohen's $\kappa$ at lag $h$ for a Markov chain $Y_t$ with state space $\mathcal{S}=\{1,2,.\ldots,q\}$. $Y_t$ is assumed to follow an AR-OSM($2$). In the following calculations we consider that $Y_t$ is stationary with stationary distribution $\bm{\pi}$ where, 
$$\pi_{ij}=P(Y_{t-1}=i,Y_t=j).$$ In addition we assume $Y_t$ as time-homogeneous, so the transition matrix $M$ with elements $M_{(\ell,i),(i,j)}$, 
$$M_{(\ell,i),(i,j)}=P(Y_t=j|Y_{t-1}=i,Y_{t-2}=\ell)$$
is constant over time.  Thus, for elemenst $M_{(\ell,i),(i,j)}$ that are specified based on \eqref{pk}, \eqref{p1}, it holds that:
$$P(Y_t=j|Y_{t-1}=i,Y_{t-2}=\ell)=P(Y_s=j|Y_{s-1}=i,Y_{s-2=\ell})~~ \forall t,s.$$

 For $h=1$, serial dependence is given  by: 

    \begin{align*}
        \kappa_o(1)&=\frac{\sum\limits_{i=1}^{q-1}(f^{(1)}_{ii}-f^{2}_i)}{\sum\limits_{i=1}^{q}f_i(1-f_i)}\\
        &=\frac{\sum\limits_{i=1}^{q-1} \Big[\Big( \sum\limits_{j_1=1}^{i}\sum\limits_{j_2=1}^{i} P(Y_t=j_2,Y_{t-1}=j_1)\Big)- (\sum\limits_{j=1}^{i}P(Y_t=j))^2\Big]}{\sum\limits_{i=1}^{q-1} \Big[(\sum\limits_{j=1}^{i}P(Y_t=j))(1-\sum\limits_{j=1}^{i}P(Y_t=j))\Big]}\\
       &=\frac{\sum\limits_{i=1}^{q-1} \Big[\Big( \sum\limits_{j_1=1}^{i}\sum\limits_{j_2=1}^{i}\pi_{j_1j_2} \Big)- \Big(\sum\limits_{j=1}^{i} \Big( \sum\limits_{\ell \in \mathcal{S}} \pi_{\ell j}\Big)\Big)^2\Big]}{\sum\limits_{i=1}^{q-1} \Big[  \Big(\sum\limits_{j=1}^{i} \Big( \sum\limits_{\ell \in \mathcal{S}} \pi_{\ell j}\Big)\Big) \Big(1-\sum_{j=1}^{i} \Big( \sum\limits_{\ell \in \mathcal{S}} \pi_{\ell j}\Big)\Big) \Big]}.
    \end{align*}

    For $h=2$, ordinal Cohen's $\kappa$ for AR-OSM($2$) is expressed by:
 \begin{align*}
        \kappa_o(2)&=\frac{\sum\limits_{i=1}^{q-1}(f^{(2)}_{ii}-f^{2}_i)}{\sum\limits_{i=1}^{q}f_i(1-f_i)}\\
        &=\frac{\sum\limits_{i=1}^{q-1} \Big[\Big( \sum\limits_{j_1=1}^{i}\sum\limits_{j_2=1}^{i} P(Y_t=j_2,Y_{t-2}=j_1)\Big)- (\sum\limits_{j=1}^{i}P(Y_t=j))^2\Big]}{\sum\limits_{i=1}^{q-1} \Big[(\sum\limits_{j=1}^{i}P(Y_t=j))(1-\sum\limits_{j=1}^{i}P(Y_t=j))\Big]}\\
       &=\frac{\sum\limits_{i=1}^{q-1} \Big[\Big( \sum\limits_{j_1=1}^{i}\sum\limits_{j_2=1}^{i} \Big(\sum\limits_{\ell \in \mathcal{S}} P(Y_t=j_2|Y_{t-1}=\ell,Y_{t-2}=j_1)P(Y_{t-1}=\ell,Y_{t-2}=j_1) \Big) \Big)- \Big(\sum\limits_{j=1}^{i} \Big( \sum\limits_{\ell \in \mathcal{S}} \pi_{\ell j}\Big)\Big)^2\Big]}{\sum\limits_{i=1}^{q-1} \Big[  \Big(\sum\limits_{j=1}^{i} \Big( \sum\limits_{\ell \in \mathcal{S}} \pi_{\ell j}\Big)\Big) \Big(1-\sum\limits_{j=1}^{i} \Big( \sum\limits_{\ell \in \mathcal{S}} \pi_{\ell j}\Big)\Big) \Big]}\\
       &=\frac{\sum\limits_{i=1}^{q-1} \Big[\Big( \sum\limits_{j_1=1}^{i}\sum\limits_{j_2=1}^{i} \Big(\sum\limits_{\ell \in \mathcal{S}}M_{(j1,\ell),(\ell,j_2)}\pi_{j_1 \ell} \Big) \Big)- \Big(\sum\limits_{j=1}^{i} \Big( \sum\limits_{\ell \in \mathcal{S}} \pi_{\ell j}\Big)\Big)^2\Big]}{\sum\limits_{i=1}^{q-1} \Big[  \Big(\sum\limits_{j=1}^{i} \Big( \sum\limits_{\ell \in \mathcal{S}} \pi_{\ell j}\Big)\Big) \Big(1-\sum\limits_{j=1}^{i} \Big( \sum\limits_{\ell \in \mathcal{S}} \pi_{\ell j}\Big)\Big) \Big]}\\  
    \end{align*}
    while for the general case of $h>2$, we have that 
\begin{align*}
        \kappa_o&(h)=\frac{\sum\limits_{i=1}^{q-1}(f^{(h)}_{ii}-f^{2}_i)}{\sum\limits_{i=1}^{q}f_i(1-f_i)}\\
        &=\frac{\sum\limits_{i=1}^{q-1} \Big[\Big( \sum\limits_{j_1=1}^{i}\sum\limits_{j_2=1}^{i} P(Y_t=j_2,Y_{t-h}=j_1)\Big)- (\sum\limits_{j=1}^{i}P(Y_t=j))^2\Big]}{\sum\limits_{i=1}^{q-1} \Big[(\sum\limits_{j=1}^{i}P(Y_t=j))(1-\sum\limits_{j=1}^{i}P(Y_t=j))\Big]}\\
       &=\frac{\sum\limits_{i=1}^{q-1} \Big[\Big( \sum\limits_{j_1=1}^{i}\sum\limits_{j_2=1}^{i} \Big(\sum\limits_{\bm{\ell} \in \mathcal{S}} P(Y_t=j_2|Y_{t-1}=\ell_1,\ldots,Y_{t-h}=j_1)\ldots P(Y_{t-h+1}=\ell_{h-1},Y_{t-h}=j_1) \Big) \Big)- \Big(\sum\limits_{j=1}^{i} \Big( \sum\limits_{\ell \in \mathcal{S}} \pi_{\ell j}\Big)\Big)^2\Big]}{\sum\limits_{i=1}^{q-1} \Big[  \Big(\sum\limits_{j=1}^{i} \Big( \sum\limits_{\ell \in \mathcal{S}} \pi_{\ell j}\Big)\Big) \Big(1-\sum\limits_{j=1}^{i} \Big( \sum\limits_{\ell \in \mathcal{S}} \pi_{\ell j}\Big)\Big) \Big]}\\
       &=\frac{\sum\limits_{i=1}^{q-1} \Big[\Big( \sum\limits_{j_1=1}^{i}\sum\limits_{j_2=1}^{i} \Big( \sum\limits_{\bm{\ell} \in \mathcal{S}} M_{(\ell_2,\ell_1),(\ell_1,j_2)}M_{(\ell_3,\ell_2),(\ell_2,\ell_1)}\ldots M_{(\ell_{h-2},\ell_{h-1}),(\ell_{h-1},j_1)}\pi_{j_1 \ell_{h-1}} \Big) \Big)- \Big(\sum\limits_{j=1}^{i} \Big( \sum\limits_{\ell \in \mathcal{S}} \pi_{\ell j}\Big)\Big)^2\Big]}{\sum\limits_{i=1}^{q-1} \Big[  \Big(\sum\limits_{j=1}^{i} \Big( \sum\limits_{\ell \in \mathcal{S}} \pi_{\ell j}\Big)\Big) \Big(1-\sum\limits_{j=1}^{i} \Big( \sum\limits_{\ell \in \mathcal{S}} \pi_{\ell j}\Big)\Big) \Big]}\\  
    \end{align*}
In the above ${\bm \ell} = (\ell_1,\ldots,\ell_{h-1})$ with all $\ell_j \in \cal S$ so $\sum\limits_{\bm \ell}$ implies a multiple summation.

\section{Simulation study}
\label{Sec4}

In this section, we provide two simulation studies. In the first case, we would like to check the performance of estimators under different sample sizes. In the second case, we would like to check how different values of $\phi$'s and $\beta$'s affects serial dependence of the series.

\subsection{Case 1: Evaluate performance of estimators}
We assume two AR-OSMs: one of order $1$ and one of order $2$. From each model we have simulated $B=500$ samples of sample sizes $T=100, 500, 1000$. 

\begin{itemize}
    \item AR-OSM($1$)\\
We consider the following autoregressive ordered stereotype model of order $1$ for an ordinal process $Y_t$, $t=1\ldots,T$ with state space $\mathcal{S}=\{1,2,3,4\}$:
\begin{align*}
    \log{\left(\frac{P(Y_t=k|\mathcal{F}_{t-1})}{P(Y_t=1|\mathcal{F}_{t-1})}\right)}=\alpha_k+\phi_k\beta Y_{t-1}, ~~k=2,\ldots,q
\end{align*}
where, $\mathcal{F}_{t-1}=\bm{\sigma}\{Y_{i},i\leq t-1\}$, $\{\alpha_1=0,\alpha_2=0.25,\alpha_3=0.10,\alpha_4=0.15\}$, $\beta=0.25$ and $\{\phi_1=0,\phi_2=0.2,\phi_3=0.5,\phi_4=1\}$.
The results are presented in Figure \ref{fig.Sim1}. 

\begin{figure}[H]
\begin{center}
\includegraphics[scale=0.45]{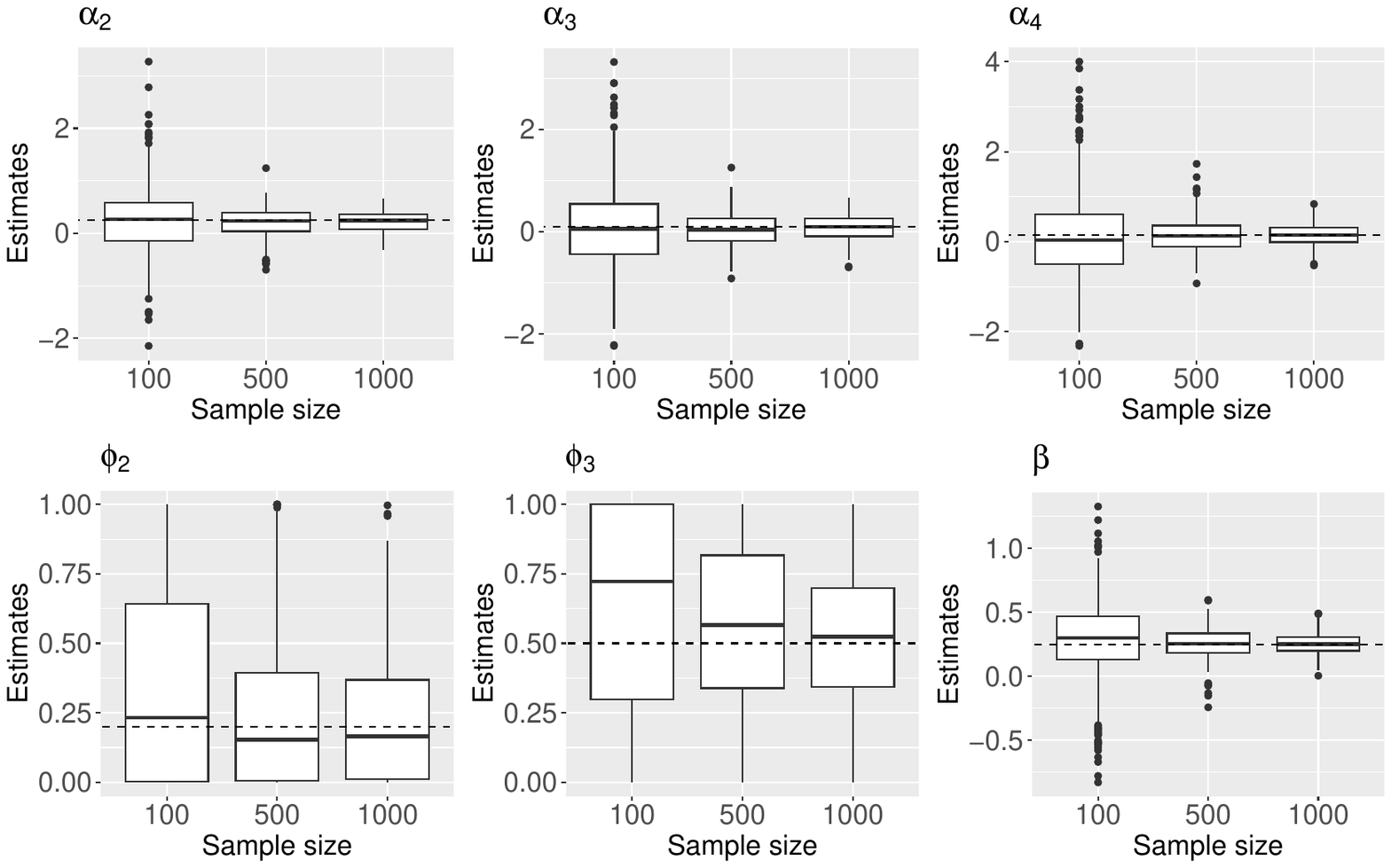}\caption{\label{fig.Sim1}  Boxplots of estimated parameters for different
sample sizes. Dashed line is the true value of the parameter}
\end{center}
\end{figure}

    \item AR-OSM($2$)\\
We consider the following autoregressive ordered stereotype model of order $2$ for an ordinal process $Y_t$, $t=1\ldots,T$ with state space $\mathcal{S}=\{1,2,3,4\}$:
\begin{align*}
    \log{\left(\frac{P(Y_t=k|\mathcal{F}_{t-1})}{P(Y_t=1|\mathcal{F}_{t-1})}\right)}=\alpha_k+\phi_k\beta_1 Y_{t-1}+\phi_k\beta_2 Y_{t-2}, ~~k=2,\ldots,q
\end{align*}
where, $\mathcal{F}_{t-1}=\bm{\sigma}\{Y_{i},i\leq t-1\}$, $\{\alpha_1=0,\alpha_2=0.25,\alpha_3=0.10,\alpha_4=0.15\}$, $\beta_1=0.30$, $\beta_2=0.15$ and $\{\phi_1=0,\phi_2=0.2,\phi_3=0.5,\phi_4=1\}$. 
The results are presented in Figure \ref{fig.Sim2}.

\begin{figure}[H]
\begin{center}
\includegraphics[scale=0.45]{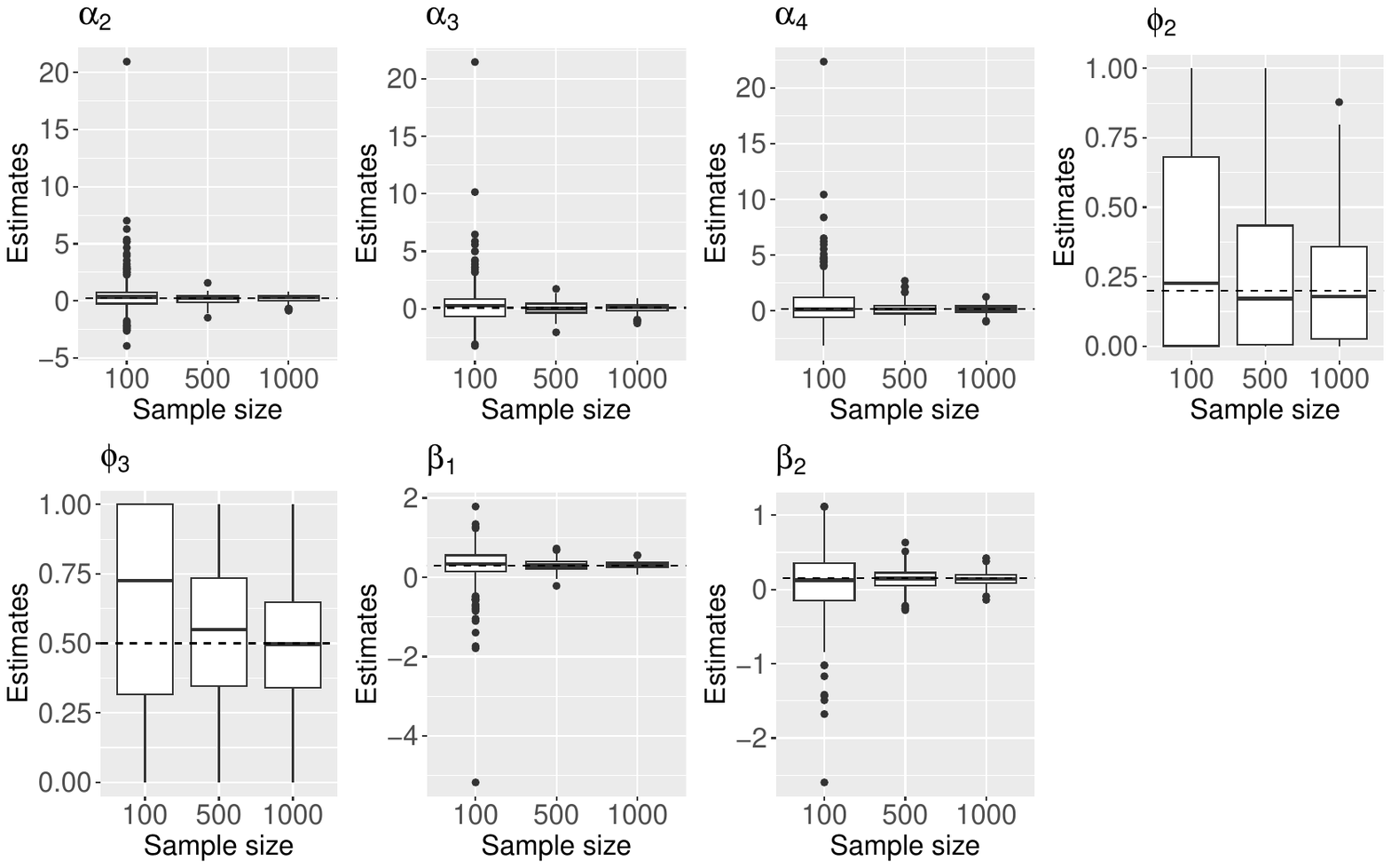}\caption{\label{fig.Sim2}  Boxplots of estimated parameters for different
sample sizes. Dashed line is the true value of the parameter}
\end{center}
\end{figure}
\end{itemize}

Based on the results from both cases, Figures \ref{fig.Sim1},\ref{fig.Sim2}, we conclude that as sample size increases variance decreases, implying efficient estimators. As bias is concerned, it seems that for small sample size ($T=100$)  there is a small bias for estimators of $\phi$'S parameters. However, for larger sample sizes the bias gradually disappears. 

\subsection{Case 2: Parameters effect on serial dependence}
At this point we would like to examine how different values of $\phi$'s and $\beta$'s parameters  affects the serial dependence of the process. For this reason, we assume an AR-OSM of order $1$, with $\alpha_1=0,\alpha_2=1,\alpha_3=4, \alpha_4=0.5$. As score parameters are concerned, we consider three scenarios:
\begin{table}[H]
    \centering
    \begin{tabular}{|c|cccc|}
    \hline
        & $\phi_1$ & $\phi_2$ & $\phi_3$ & $\phi_4$\\
        \hline
         Scenario 1&  0 & 0.33 & 0.66 & 1\\
         Scenario 2&  0 & 0.5  & 0.7 & 1\\
         Scenario 3&  0 & 0.3  & 0.5 & 1\\
         \hline
    \end{tabular}
    \caption{Score parameters for each scenario \label{tab.scen}}
\end{table}
while for each scenario we take a grid of values for $\beta_1$ from $-5 $ to $5$ with step $0.5$. 

\begin{figure}[H]
    \centering
    \includegraphics[width=0.7\linewidth]{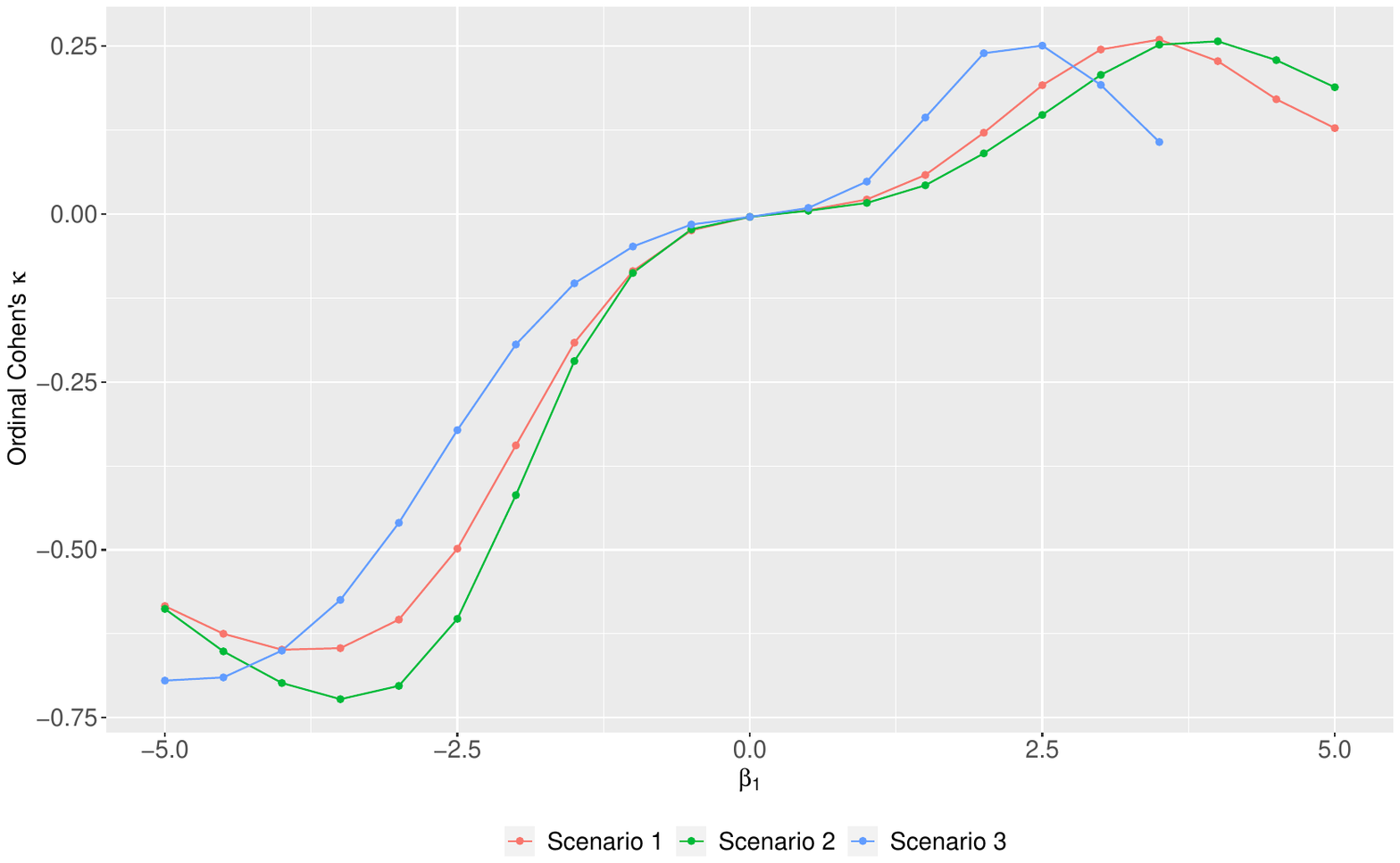}
    \caption{ Mean ordinal Cohen's $\kappa$ per scenario and $\beta_1$ values based on $B=500$ simulations\label{fig:sim_cohen}
}
\end{figure}

From each case we have simulated $B=500$ samples of sample size $T=500$. For each sample we estimated the ordinal Cohen's $\kappa$ at lag $1$ and finally we present their mean values in Figure \ref{fig:sim_cohen}. We conclude that sign of $\beta_1$ affects the type of serial dependence: positive $\beta_1$ lead to positive serial dependence, while negative $\beta_1$ conclude negative serial dependence. The relation between $\beta_1$ and serial dependence is not linear, so it doesn't mean that as $|\beta_1|$ increases so does the serial dependence. However, for $\beta_1$ close to $0$ the serial dependence is weak as it is expected.  

\noindent
\textbf{Remark}: For Scenario 3, ordinal Cohen's $\kappa$ is not available for $\beta_1=4,4.5,5$, while process stays in one state. For positive $\beta_1$ previous higher states increases the probability of staying in higher states. The magnitude of $\beta_1$ controls the strength of this shift. 

To examine how magnitude and sign of $\beta_1$ affects the results, we
present the theoretical transition matrices under models with different parameters. We consider the autoregressive model of order $1$ with $\alpha$ parameters as they have been defined. For score parameters we assume the three scenarios presented in Table \ref{tab.scen}, while for $\beta_1$ coefficients we assume three values $\beta_1=-3.5, 1,3.5$. For each case we present the transition matrix at lag $1$, denoted by $P$ with elements $P_{ij}=P(Y_t=j|Y_{t-1}=i), ~i,j=1,\ldots,4$. 

The results are presented in Tables \ref{tab. P1}-\ref{tab. P3}. For all scenarios we can see that for negative $\beta_1$, for high previous states, the probabilities for lower states are greater than for a current high state. On the other hand, for positive $\beta_1$, high previous state leads to hight probabilities for higher states, while magnitude of $\beta_1$ plays an essential role. High positive $\beta_1$ leads to mass concentration on high states.  

\begin{itemize}
    \item Scenario 1
\begin{table}[H]
\centering
\renewcommand{\arraystretch}{1.2}
\begin{tabular}{ccc}
$\beta_1=-3.5$ & $\beta_1=1$ & $\beta_1=3.5$ \\[6pt]

P=$\begin{bmatrix}
0.137 & 0.117 & 0.740 & 0.007\\
0.553 & 0.149 & 0.297 & 0.001\\
0.878 & 0.075 & 0.047 & \approx0\\
0.969 & 0.026 & 0.005 & \approx0\\
\end{bmatrix}$
&
P=$\begin{bmatrix}
0.009 & 0.033 & 0.919 & 0.039\\
0.004 & 0.024 & 0.917 & 0.055\\
0.002 & 0.017 & 0.905 & 0.076\\
0.001 & 0.012 & 0.883 & 0.104\\
\end{bmatrix}$
&
P=$\begin{bmatrix}
0.002 & 0.014 & 0.895 & 0.089\\
\approx0 & 0.004 & 0.751 & 0.245\\
\approx0 & 0.001 & 0.482 & 0.517\\
\approx0 & \approx0 & 0.221 & 0.779\\

\end{bmatrix}$
\end{tabular}
\caption{Transition matrices for different values of $\beta_1$ for Scenario 1 \label{tab. P1}}
\end{table}

    \item Scenario 2 
    
\begin{table}[H]
\centering
\renewcommand{\arraystretch}{1.2}
\begin{tabular}{ccc}
$\beta_1=-3.5$ & $\beta_1=1$ & $\beta_1=3.5$ \\[6pt]

P=$\begin{bmatrix}
0.160 & 0.076 & 0.756 & 0.008\\
0.671 & 0.055 & 0.273 & 0.001\\
0.953 & 0.014 & 0.033 & \approx0\\
0.995 & 0.002 & 0.003 & \approx0\\
\end{bmatrix}$
&
P=$\begin{bmatrix}
0.008 & 0.037 & 0.917 & 0.037\\
0.004 & 0.031 & 0.915 & 0.050\\
0.002 & 0.025 & 0.906 & 0.067\\
0.001 & 0.020 & 0.890 & 0.089\\
\end{bmatrix}$
&
P=$\begin{bmatrix}
0.001 & 0.022 & 0.899 & 0.078\\
\approx0 & 0.010 & 0.794 & 0.196\\
\approx0 & 0.004 & 0.585 & 0.412\\
\approx0 & 0.001 & 0.331 & 0.668\\
\end{bmatrix}$
\end{tabular}
\caption{Transition matrices for different values of $\beta_1$ for Scenario 2 \label{tab. P2}}
\end{table}

\item Scenario 3 
    
\begin{table}[H]
\centering
\renewcommand{\arraystretch}{1.2}
\begin{tabular}{ccc}
$\beta_1=-3.5$ & $\beta_1=1$ & $\beta_1=3.5$ \\[6pt]

P=$\begin{bmatrix}
0.087 & 0.083 & 0.826 & 0.004\\
0.335 & 0.112 & 0.553 & 0.001\\
0.713 & 0.083 & 0.204 & \approx0\\
0.917 & 0.037 & 0.046 & \approx0\\
\end{bmatrix}$
&
P=$\begin{bmatrix}
0.010 & 0.037 & 0.908 & 0.045\\
0.006 & 0.030 & 0.891 & 0.073\\
0.004 & 0.023 & 0.857 & 0.116\\
0.002 & 0.018 & 0.801 & 0.179\\
\end{bmatrix}$
&
P=$\begin{bmatrix}
0.003 & 0.021 & 0.832 & 0.145\\
\approx0 & 0.006 & 0.497 & 0.497\\
\approx0 & 0.001 & 0.148 & 0.851\\
\approx0 & \approx0 & 0.029 & 0.971\\
\end{bmatrix}$
\end{tabular}
\caption{Transition matrices for different values of $\beta_1$ for Scenario 3 \label{tab. P3}}
\end{table}
\end{itemize}

\section{Application: Sleep data}
\label{Sec5}
\subsection{About data}
\begin{figure}[H]
     \centering
     \includegraphics[scale=0.40]{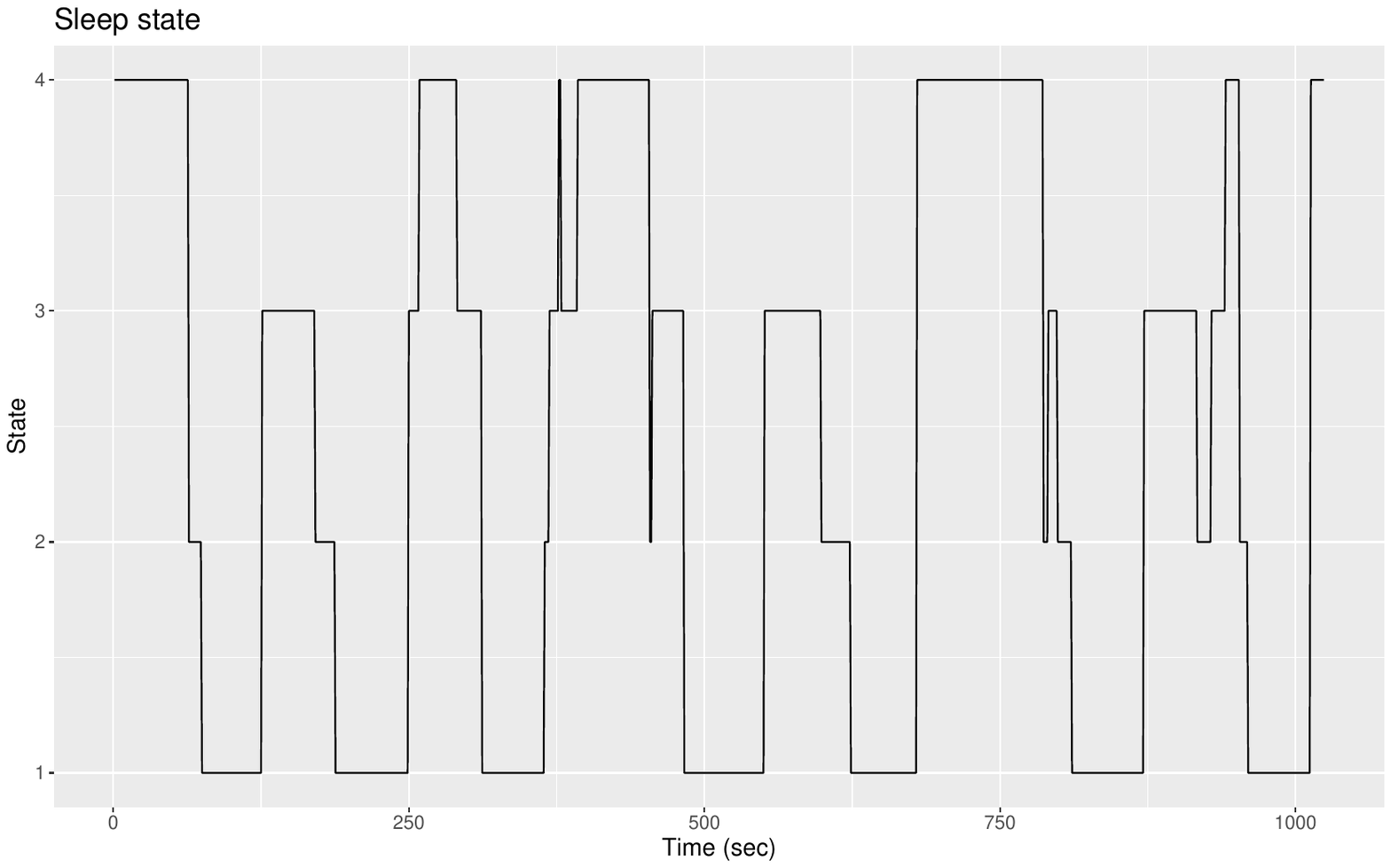}
     \caption{Sleep state of the newborn infant per 30 seconds 
 \label{fig:Sleep_state}}
 \end{figure}

We have $T=1024$ sleep state measurements and the heart rate $R_t$ of a newborn infant every $30$ seconds. The data are available in $\texttt{R}$ package \texttt{weightedCL}.  The sleep states are classified as: $(4)$ awake, $(1)$ quiet sleep, $(2)$ indeterminate sleep, $(3)$ active sleep. The time series data are presented in Figure \ref{fig:Sleep_state}. 

The data can also be seen as an ordinal time series $Y_t$, where the states $\mathcal{S}=\{(4),(1),(2),(3)\}$ present the following ordering: $(4) < (1) < (2) < (3)$. The ordinal nature of the process is important for modeling. Treating ordinal time series as nominal ignores the natural ordering of the states that provides an additional information about transitions and  serial dependence. In addition, in contrast to nominal processes, distance between states is meaningful due to the rank among states. For example, a transition between states can be interpreted as an improvement or deterioration rather than simply a change in category. In our case, the ordering of sleep states reflects increasing depth of sleep rather than wakefulness.     

\subsection{Measuring serial dependence}
Before fitting the model it is useful examining whether there is serial dependence. As we are working with ordinal time series we prefer using a serial dependence measure that takes into consideration the nature of the data, like the ordinal Cohen's $\kappa$ that has already been mentioned. We also provide its partial version that is helpful on model selection. The estimates are presented in Table \ref{tab:Ordinal_cohen}. 

\begin{table}
\begin{footnotesize}
    \centering
    \begin{tabular}{|c|cccccccccc|}
    \hline
    lag & 1 & 2 & 3 & 4 & 5 & 6 & 7 & 8 & 9 & 10 \\
    \hline
     Ordinal Cohen's $\kappa$ & 0.956 & 0.912 & 0.873 & 0.833 & 0.794 & 0.754 & 0.714 & 0.676 & 0.642 & 0.610\\
     \hline
    Partial Ordinal Cohen's $\kappa$ & 0.956 & -0.024 & 0.033 & -0.022 & -0.019 & -0.023 & -0.023 & -0.005 & 0.014 & 0.018\\
\hline

    \end{tabular}
    \end{footnotesize}
    \caption{Ordinal Cohen's $\kappa$ and its partial version for different lags\label{tab:Ordinal_cohen}}    
\end{table}

Based on Table \ref{tab:Ordinal_cohen}, ordinal Cohen's $\kappa$ shows a gradual decay across increasing lags, indicating the presence of serial dependence. Its partial version exhibits a clear cut-off after lag $1$. It is large at lag $1$  and remains close to zero for all higher lags, with small oscillations around zero that are consistent with random variation. This pattern indicates that an autoregressive term of order $1$ is adequate.

\subsection{Model selection and Estimation}
We try different models to find the optimal one. A summary of them is presented in Table \ref{tab.summary models}, while a more detailed presentation follows.  

Considering the results of Table \ref{tab:Ordinal_cohen}, we start with an AR-OSM model of order $1$ including the heart rate (Model 1). 
    \begin{align*}
    \log{\left(\frac{P(Y_t=k|\mathcal{F}_{t-1},R_t)}{P(Y_t=(4)|\mathcal{F}_{t-1},R  _t)}\right)}=\alpha_k+\phi_k\beta_1Y_{t-1}+\phi_k\gamma R_t, ~~k=(1),(2),(3)
\end{align*}
where $\alpha_{(4)}=0$ and $0=\phi_{(4)}\leq\phi_{(1)}\leq \phi_{(2)}\leq \phi_{(3)}=1$.

Then, we fit a multinomial logit model with an autoregressive term of order $1$ \citep{fokianos2003regression} and the heart rate as covariate (Model 2). 

    \begin{align*}
    \log{\left(\frac{P(Y_t=k|\mathcal{F}_{t-1},R_t)}{P(Y_t=(4)|\mathcal{F}_{t-1},R_t)}\right)}=\alpha_k+\beta_{1k}Y_{t-1}+\gamma_k R_t, ~~k=(1),(2),(3)
\end{align*}

Comparing Models $1$ and $2$ we examine the adequacy of the ordinal trend. The two models are nested, as it holds that they are equivalent in case of $\phi_k\beta_1=\beta_{1k}$ and $\phi_k\gamma=\gamma_k$, for $k=(1),(2)$ with $0=\phi_{(4)}\leq\phi_{(1)}\leq \phi_{(2)}\leq \phi_{(3)}=1$.
Based on a Likelihood Ratio Test (LRT) we conclude that Model 1 is better (p-value=$0.269>0.05$). 

At this point, we would like to examine whether it is essential to keep heart rate in the model. Thus, we fit an AR-OSM model of order $1$ without heart rate (Model 3):
    \begin{align*}
    \log{\left(\frac{P(Y_t=k|\mathcal{F}_{t-1})}{P(Y_t=(4)|\mathcal{F}_{t-1})}\right)}=\alpha_k+\phi_k\beta_1Y_{t-1}, ~~k=(1),(2),(3)
\end{align*}
where $\alpha_{(4)}=0$ and $0=\phi_{(4)}\leq\phi_{(1)}\leq \phi_{(2)}\leq \phi_{(3)}=1$. Also using LRT we conclude that heart rate doesn't improve the model (p-value=$0.089>0.05$), so Model 3 is better than Model 1. 

It is also interesting checking whether the autoregressive term of order $1$ is adequate to describe the data or a higher order of autoregressive terms should be considered. So, we fit an AR-OSM of order $2$ (Model 4): 
    \begin{align*}
    \log{\left(\frac{P(Y_t=k|\mathcal{F}_{t-1})}{P(Y_t=(4)|\mathcal{F}_{t-1})}\right)}=\alpha_k+\phi_k\beta_1Y_{t-1}+\phi_k\beta_2Y_{t-2}, ~~k=(1),(2),(3)
\end{align*}
where $\alpha_{(4)}=0$ and $0=\phi_{(4)}\leq\phi_{(1)}\leq \phi_{(2)}\leq \phi_{(3)}=1$. Based on LRT  between Models 3 and 4 (pvalue=$0.46>0.05$), Model 3 is adequate as it is expected from the results of the partial Cohen's $\kappa$. 

Finally, we check whether we can have a more parsimonious version of the optimal Model 3, by fitting its counterpart with the proportionality assumption (Model 5). This model assumes proportional slopes:
   \begin{align*}
    \log{\left(\frac{P(Y_t=k|\mathcal{F}_{t-1})}{P(Y_t=(4)|\mathcal{F}_{t-1})}\right)}=\alpha_k+\phi_k\beta_1Y_{t-1}, ~~k=(1),(2),(3)
\end{align*}
where $\alpha_{(4)}=0$ and $\phi_{(4)}=0, \phi_{(1)}=0.33, \phi_{(2)}=0.66, \phi_{(3)}=1$.
In fact, it can be seed as an autoregressive version (of order $1$) of the proportional odds adjacent categories-logit model. 
Comparing Model 5 and Model 3 we can conclude whether states are equidistant and then whether the effect of previous states is the same across outcomes. The comparison based on LRT indicates that Model 3 is better (p-value$<0.001$).   

\begin{table}[H]
    \centering
    \begin{tabular}{cccc}
    \hline
    Model & Log-lik & df & BIC \\
    \hline
         Model 1 &  -359.4111 & 7 & 767.34\\
         Model 2 &  -358.0973 & 9 & 778.57\\
         Model 3 & -360.8588  & 6 & 763.30\\
         Model 4 & -360.5868  & 7 & 769.69\\
         Model 5 & -380.0833 & 4 & 787.89\\
         \hline
    \end{tabular}
    \caption{Log-likelihood, number of parameters and BIC for the different models  \label{tab.summary models}}
    \label{tab:placeholder}
\end{table}

\begin{comment}
\begin{table}[H]
    \centering
    \begin{tabular}{cccc}
    \hline
        & Model 1 & Model 2 & Model 3 \\   
    \hline
    $\hat{\alpha}_{(1)}$ & $\underset{(0.609)}{-9.219}$ &$\underset{(0.608)}{-9.198}$ & $\underset{0.332}{-6.883}$\\
    $\hat{\alpha}_{(2)}$ & $\underset{(1.002)}{-19.322}$ & $\underset{(1.002)}{-19.288}$ & $\underset{(0.830)}{-19.462}$\\
    $\hat{\alpha}_{(3)}$ & $\underset{(1.701)}{-31.352}$ & $\underset{(1.702)}{-31.321}$ & $\underset{(1.569)}{-35.806}$\\
    $\hat{\phi}_{(1)}$ & $\underset{(0.023)}{0.461}$ \tablefootnote{Similar results with \texttt{OSM()}} & $\underset{(0.023)}{0.461}$ & -  \tablefootnote{In Model 3, scores $\phi_k$ are fixed and equally spaced: 0.33 and 0.66.} \\
    $\hat{\phi}_{(2)}$ & $\underset{(0.021)}{0.731}$ & $\underset{(0.021)}{0.731}$ & - \\
    $\hat{\beta}_1$ & $\underset{(0.642)}{13.538}$ & $\underset{(1.137)}{14.133}$ & $\underset{(0.611)}{14.238}$ \\
    $\hat{\beta}_2$ &  - & $\underset{(0.939)}{-0.610}$ & - \\
       \hline 
       Log-lik &  -360.8588 & -360.5868 & - 380.0833  \\
       \hline 
    \end{tabular}
    \caption{Estimates with standard errors in the parenthesis and log-likelihood of each model \label{tab:Estimates}}  
\end{table}
\end{comment}

\begin{table}[H]
    \centering
    \begin{tabular}{cc}
    \hline
   Coefficient & $\underset{(se)}{\mbox{Estimate}}$\\
   \hline
  $\hat{\alpha}_{(1)}$ & $\underset{(0.609)}{-9.219}$ \\
  $\hat{\alpha}_{(2)}$ & $\underset{(1.002)}{-19.322}$\\
  $\hat{\alpha}_{(3)}$ & $\underset{(1.701)}{-31.352}$\\
  $\hat{\phi}_{(1)}$ & $\underset{(0.023)}{0.461}$ \\
  $\hat{\phi}_{(2)}$ & $\underset{(0.021)}{0.731}$\\
  $\hat{\beta}_1$ & $\underset{(0.642)}{13.538}$\\
  \hline
    \end{tabular}
    \caption{Estimates and standard error  for the optimal Model 3}
    \label{tab:placeholder}
\end{table}

\begin{table}[H]
    \centering
    \begin{tabular}{|c|c|}
    \hline
    States & Distance\\
    \hline
      $\phi_{(4)}-\hat{\phi}_{(1)}$\rule{0pt}{3ex} & 0.461 \\
      $\hat{\phi}_{(1)}-\hat{\phi}_{(2)}$\rule{0pt}{3ex} & 0.27\\
      $\hat{\phi}_{(2)}-\phi_{(3)}$\rule{0pt}{3ex} & 0.269\\
      \hline
    \end{tabular}
    \caption{Distance between states based on Model 3 \label{tab:Distance_states}} 
\end{table}

\begin{figure}[H]
\begin{center}
    \includegraphics[scale=0.5]{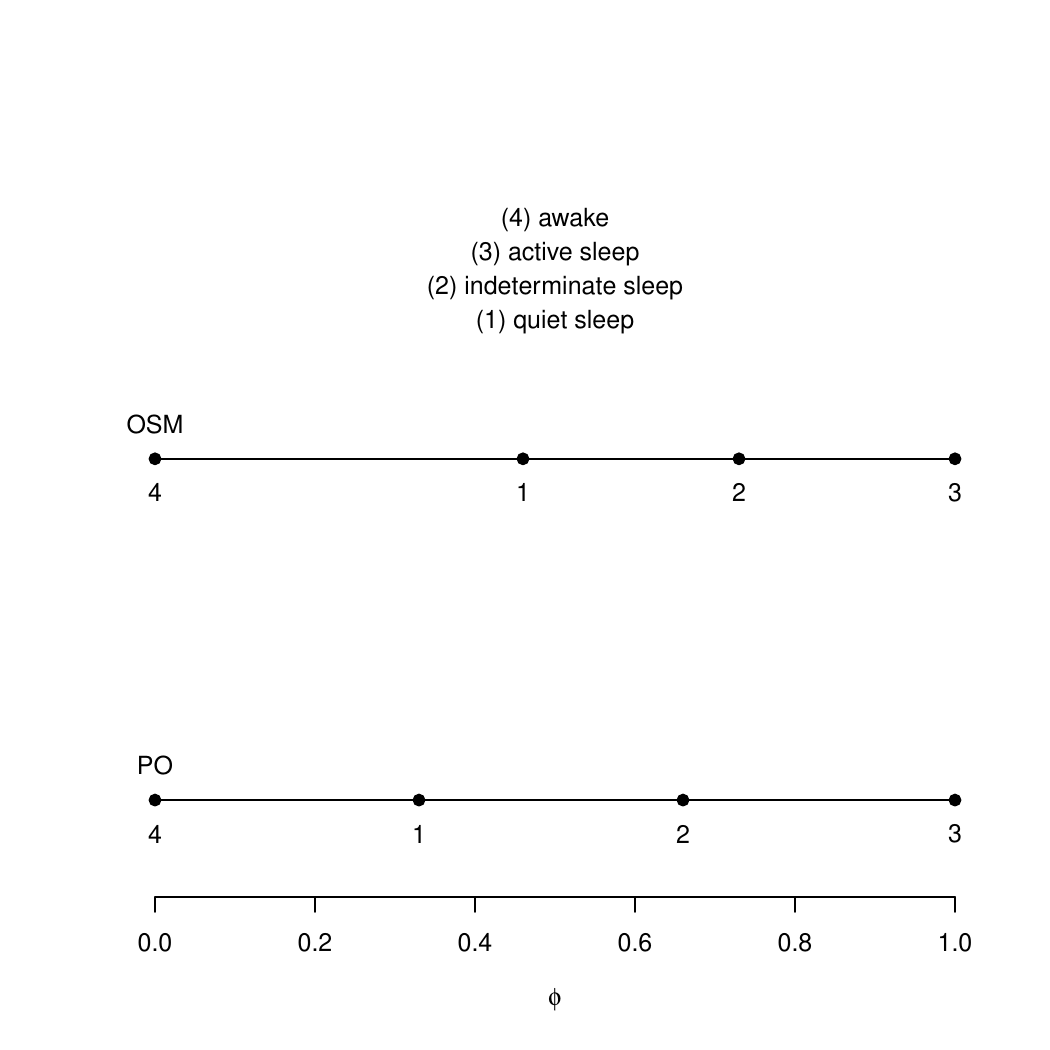}
    \caption{\label{scale} The estimated $\phi$'s based on the OSM and the proportional models. There is evidence against the proportionality }
    \end{center}
\end{figure}

\begin{figure}[H]
\begin{center}
    \includegraphics[scale=0.5]{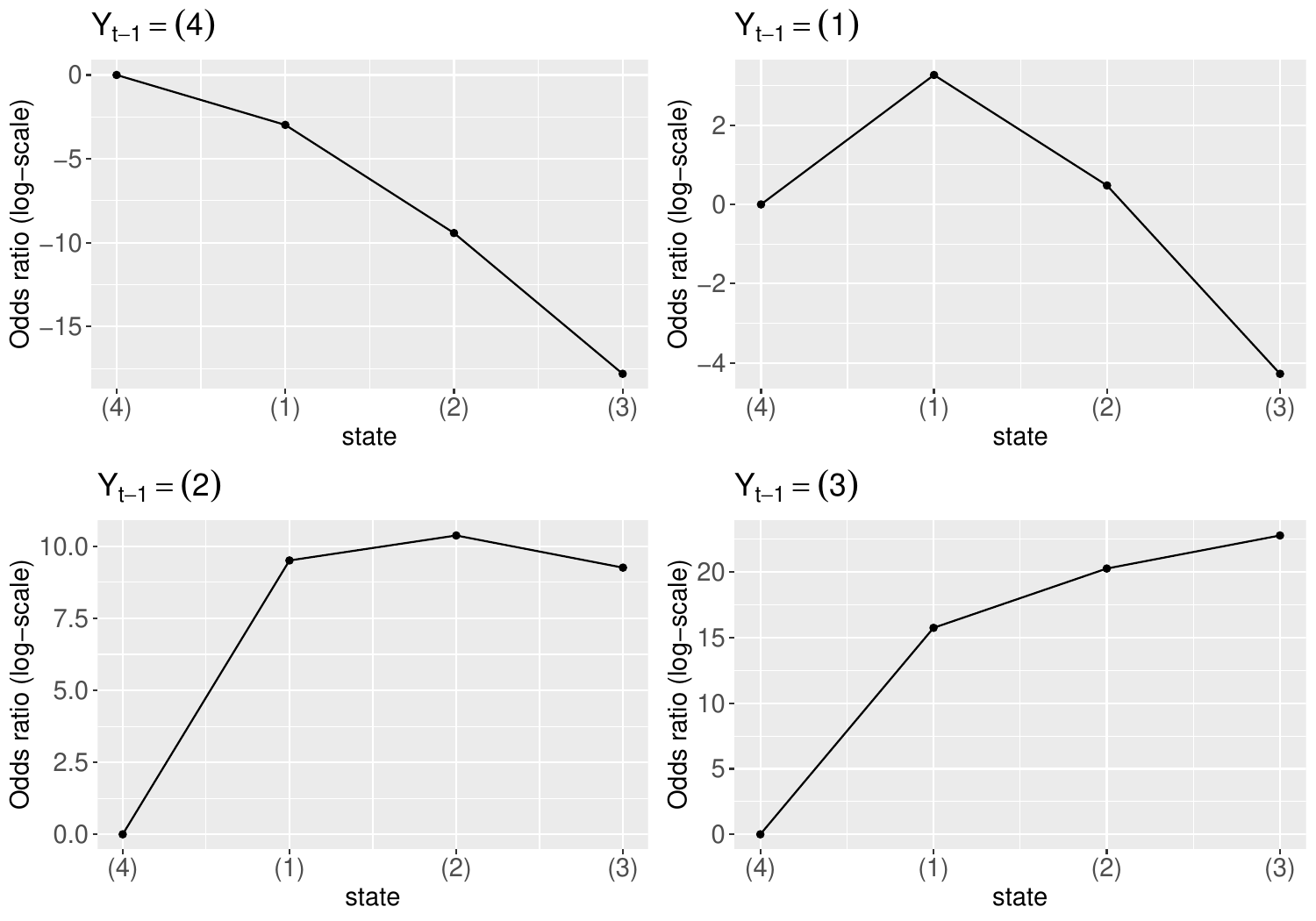}
    \caption{\label{log_odds} Odds-ratios in log scale for each ordinal state conditional on the  previous state }
    \end{center}
\end{figure}

Figure \ref{scale} shows the estimated $\phi$'s with the OSM and the proportional model. For the proportional model we assume that they are equidistant. When we estimate those distance via the OSM model we see that this is not the case. In addition, Table \ref{tab:Distance_states}, presents the estimated distance between states under Model 3. We observe that the latent distance between states $(4)$ and $(1)$ is substantially larger than the distances between $(1)-(2)$ and $(2)-(3)$, which are similar. This suggests unequal spacing of the ordinal states, supporting Model 3 over Model 5. In particular, the transition from “awake” to “quiet sleep” appears more difficult than transitions among intermediate sleep stages. In addition, as $\hat{\beta}_1$ of Model 3 is concerned, we notice that is large and positive. This means that the higher the previous sleep state the higher the probability of  of remaining in a higher current state.
Finally, Figure \ref{log_odds} shows the  odds-ratios in log scale for each ordinal state conditional on the previous state previous state. The pattern is distinct for each state. For instance, when $Y_{t-1}=(4)$ (awake), the log-odds of remaining in that state (awake) are the largest, while the log-odds of transitioning to quieter states become increasingly negative, reflecting a strong 
tendency to persist in wakefulness. As the previous state decreases toward quiet sleep $(1)$, this pattern progressively reverses, with 
higher log-odds now associated with the lower sleep states.

\subsection{Forecasting}
At this point we would like to use the final optimal model (Model 3) for forecasting. For this reason, we assume that we have available data from the period $t=1,\ldots,1014$ and we would like to forecast sleep state of the newborn for the last 5 minutes ($10-$steps ahead forecasting). We fit the model based on the available information of the first $1014$ observations. Then, to forecast $Y_{t+h}$ for $h=1,\ldots,10$, we use the following probabilities for each state, conditional to the previous state: 
\begin{align*}
    P(Y_{t+h}=k|Y_{t+h-1})=\frac{\exp(\alpha_j+\phi_j\beta_1 Y_{t+h-1})}{\sum_{\ell=1}^{4}\exp(\alpha_\ell+\phi_\ell\beta Y_{t+h-1})}, ~ \mbox{for}~~k=(4),(1),(2),(3).
\end{align*}
Then, the forecast $\hat{Y}_{t+h}$ will occur as a random draw from the conditional distribution $Y_{t+h}|Y_{t+h-1}$. For $h=1$, $Y_{t+h-1}$ is known. It is the last observation of the train set, so $Y_{t+h-1}=Y_{1014}$. However, for the next steps $h>1$, we need to use the last forecast in each case: $Y_{t+h-1}=\hat{Y}_{t+h-1}$. We repeat this procedure for $B=1000$ times to approximate the predictive distribution. Then, as final forecast for each $h-$step we consider the state with the highest frequency. For comparison, we follow the same procedure to provide forecasts based on Model 5.
The results are presented in Tables \ref{Tab.Forc1} and \ref{Tab.Forc3}. As we can see, in each step ahead Model 3 predicts the correct state with higher probability than Model 5. In addition, it is noteworthy that, due to the multi-step forecasting framework, both models predict the correct state with greater confidence in the near future. As the forecast horizon increases, the predictive distribution becomes more dispersed, thereby reducing predictive certainty.

\begin{table}[H] 
\centering
\begin{tabular}{|c|cccc|c|c|} 
\hline 
  & & Relative Frequency & & & $\hat{Y}_{t+h}$\rule{0pt}{3ex} & True \\
 \hline
$h$ & (4) & (1) & (2) & (3) & & \\ 
\hline
1 & 0.939  & 0.061 & 0 & 0 & (4) & (4)\\
2 & 0.905 & 0.092 & 0.003 & 0 & (4) & (4)\\
3 & 0.868  & 0.123 & 0.007 & 0.002 & (4) & (4)\\
4 & 0.830 & 0.156 & 0.011 & 0.003 & (4) & (4)\\
5 & 0.807 & 0.174 & 0.013 & 0.006 &(4) & (4)\\
6 & 0.782 & 0.197 & 0.013 & 0.008 & (4) & (4)\\
7 &0.751 & 0.213 & 0.022 & 0.014 & (4) & (4)\\
8 & 0.715 & 0.249 & 0.019 & 0.017 & (4) & (4)\\
9 & 0.692 & 0.271 & 0.018 & 0.019 & (4) & (4)\\
10 & 0.663 & 0.287 & 0.027 & 0.023 & (4) & (4)\\
\hline
\end{tabular}
\caption{\label{Tab.Forc1} Relative frequency table of states based $1000$ simulations, final forecasts and true values from Model 3.}
\end{table}

\begin{table}[H] 
\centering
\begin{tabular}{|c|cccc|c|c|} 
\hline 
  & & Relative Frequency & & & $\hat{Y}_{t+h}$\rule{0pt}{3ex} & True \\
 \hline
$h$ & (4) & (1) & (2) & (3) & & \\ 
\hline
1 & 0.887 & 0.113 & 0 & 0 & (4) & (4)\\
2 & 0.867 & 0.129 & 0.004 & 0 & (4) & (4)\\
3 & 0.832 & 0.164 & 0.002 & 0.002 & (4) & (4)\\
4 & 0.787 & 0.202 & 0.007 & 0.004 & (4) & (4)\\
5 & 0.766 & 0.216 & 0.016 & 0.002 &(4) & (4)\\
6 & 0.742 & 0.232 & 0.019 & 0.007& (4) & (4)\\
7 & 0.722 & 0.253 & 0.013 & 0.012 & (4) & (4)\\
8 & 0.685 & 0.273 & 0.026 & 0.016 & (4) & (4)\\
9 & 0.649 & 0.310 & 0.023 & 0.018& (4) & (4)\\
10 &  0.647 & 0.313 & 0.021 & 0.019 & (4) & (4)\\
\hline
\end{tabular}
\caption{\label{Tab.Forc3} Relative frequency table of states based $1000$ simulations, final forecasts and true values from Model 5.}
\end{table}

\subsection{Stationarity}
$Y_t$ is a Markov chain of order $1$ with finite state space $\mathcal{S}=\{(4),(1),(2),(3)\}$. Let $P$ denotes the transition matrix with elements $P_{ij}=P(Y_t=j|Y_{t-1}=i)$, $i,j=(4),(1),(2),(3)$. Based on Model 3, the estimated transition matrix $\hat P_1$ is: 
\begin{align*}P_1=
\begin{bmatrix}
0.951 & 0.049 & \approx0 & \approx0\\
0.035 & 0.909 & 0.056 & \approx0\\
\approx0 & 0.240 & 0.572 & 0.187\\
\approx0 & 0.001 & 0.074 & 0.925\\  
\end{bmatrix}
\end{align*}
Based on $\hat P_1$, the Markov chain is irreducible and aperiodic, thus it is stationary. The stationary distribution, denoted by $\bm{\pi}_1=(\pi^{(1)}_{(4)},\pi^{(1)}_{(1)},\pi^{(1)}_{(2)},\pi^{(1)}_{(3)})^{'}= (0.282, 0.395,0.092,0.232)^{'}$, is given by solving the system $\bm{\pi}_1=\bm{\pi}_1P_1$, where $\sum_{\ell\in\mathcal{S}}\pi^{(1)}_{(\ell)}=1$. 

For Model 5, the transition matrix $\hat P_2$ is:
\begin{align*}P_2=
\begin{bmatrix}
0.899 & 0.101 & \approx0 & \approx0\\
0.072 & 0.891 & 0.037 & \approx0\\
\approx0 & 0.159 & 0.724 & 0.117\\
\approx0 & \approx0 & 0.046 & 0.953\\
\end{bmatrix}
\end{align*}
indicating also stationary process, with stationary distribution $\bm{\pi}_2=(\pi^{(2)}_{(4)},\pi^{(2)}_{(1)},\pi^{(2)}_{(2)},\pi^{(2)}_{(3)})^{'}=\\ (0.282, 0.395,0.092,0.232)^{'}$.  

Summarizing our findings, an autoregressive term of order $1$ seems adequate to describe the data. The AR-OSM outperforms the autoregressive proportional odds adjacent-categories logit model, as data states unequal distances between categories. More specifically, going from "awake" to "quiet sleep" is harder than transitions between deeper sleep states, which seems plausible and justify the final option. Finally, AR-OSM provides better results in forecasting compared to the proportional odds model, indicating the correct state with higher probability.   

\section{Conclusion}
\label{Sec6}
In this paper we presented an extension of the OSM in case of ordinal time series data, the AR-OSM. This regression model accounts for the serial dependence of the data by adapting autoregressive terms in its systematic component. In contrary to other autoregression models that assumes equidistant states, autoregerssive OSM is based on data to estimate the real distance between states. It respects ordinal nature of the data, while it remains parsimonious. In addition, as a regression model, it enables the incorporation of covariates, compared to other models for ordinal time series.  

Model's performance across different sample sizes was examined through a simulation study. The results showed overall good behavior but with a special focus on the bias of "core" parameters in case of small sample size. From the illustrative example on sleep state data, we confirm that in real-data applications the common assumption of equidistant states of popular models for ordinal time series may not be realistic and then the AR-OSM is more appropriate. 
Another possible extension of the OSM model in the context of ordinal time series may referred to an OSM with time varying parameters. However, the difficulty of such approach is detected in the estimation of the model due to the restriction of ``score'' parameters.  A possible approach is to assume that $\phi$ parameters change over time, while they still satisfy the ordering restrictions.

A further natural extension of the AR-OSM concerns the score parameters. In the current formulation, a single set of score parameters is shared across all covariates, implying that all predictors exhibit the same discriminant ability with respect to the ordinal outcome. The recently proposed Partial Ordered Stereotype Model \citep{egea2025partial} relaxes this assumption in the cross-sectional setting by allowing each covariate to have its own set of score parameters within a single unified framework. Extending the AR-OSM in this direction; that is, developing an autoregressive Partial Ordered Stereotype Model for ordinal time series would allow the distances between ordinal states to vary across covariates while still accounting for serial dependence, offering a considerably more flexible modelling tool. We leave this as a promising avenue for future research.

%\begin{comment}
\section*{Funding}
Daniel Fernández is a Serra-Húnter Fellow and his work has been supported by \\ MICIU/AEI/10.13039/501100011033 (Spain) and by FEDER (EU)[PID2023-148033OB-C21], by grant 2021 SGR 01421 (GRBIO) administrated by the Departament de Recerca i Universitats de la Generalitat de Catalunya (Spain), and by the Marsden Fund (Award Number E11772-5342) from New Zealand Government funding.

The research work of Anna Nalpantidi was supported by the Hellenic Foundation 
for Research and Innovation (HFRI) under the 5th Call for 
HFRI PhD Fellowships (Fellowship Number: 20535.)
%\end{comment}

\section*{Declaration of conflicting interests}
The author(s) declared no potential conflicts of interest with respect to the research, authorship, and/or publication of this article.

\bibliographystyle{apalike}
\bibliography{biblio}

\newpage

\section*{Appendix: Autocorrelation function}

While in \hyperref[Sec3]{Section 3} we derived the serial dependence using the Ordinal Cohen's kappa function, here we provide classical Pearson autocorrelation derivation assuming a numeric representation of the variable.  

\subsection{Autoregressive model of order $1$}
Let $\rho(k)$ denotes the auto-correlation function at lag $k$ for a Markov chain $Y_t$ with state space $\mathcal{S}$. In the following calculations we consider that $Y_t$ is stationary with stationary distribution $\bm{\pi}$. Under stationarity assumption it holds that $E(Y_t)=E(Y_{t-k})$. In addition we assume $Y_t$ as time-homogeneous, so in the transition matrix $P$ the elements $P_{ij}=P(Y_t=j|Y_{t-1}=i)$  
are stable, which means that they do not depend on time:
$$P(Y_t=j|Y_{t-1}=i)=P(Y_s=j|Y_{s-1}=i),~~ \forall t,s$$

$$P_{ij}^{(h)}=P(Y_t=j|Y_{t-h}=i)$$

\begin{itemize}

   \item For lag 0 we get that, 
\begin{align*}
\rho(0)&=1,\\
\gamma(0)&=E(Y_t^2)-E(Y_t)^2\\
 &=\sum_{j\in \mathcal{S}}y_t^2P(Y_t=j)-\left(\sum_{j\in\mathcal{S}}y_tP(Y_t=j)\right)^{2}.
\end{align*}

    \item For lag 1 we get that, 
\begin{align*}
    \rho(1)&=\frac{\gamma(1)}{\gamma(0)},~~ \mbox{where}\\
\gamma(1)&=\mbox{Cov}(Y_t,Y_{t-1})=E(Y_tY_{t-1})-E(Y_t)^2\\
         &=\sum_{i\in\mathcal{S}}\sum_{j\in\mathcal{S}} ijP(Y_t=j,Y_{t-1}=i)-E(Y_t)^2\\
         &=\sum_{i\in\mathcal{S}}\sum_{j\in\mathcal{S}} ij P(Y_t=j|Y_{t-1}=i)P(Y_{t-1}=i)-E(Y_t)^2\\
         &=\sum_{i \in \mathcal{S}}\sum_{j \in \mathcal{S}}ijP_{ij}\pi_i-E(Y_t)^2
\end{align*}
where conditional probabilities come from the transition matrix and marginal probabilities from the stationary distribution (due to stationarity assumption). 

\begin{comment}
    \item For lag 2 we get that:
\begin{align*}
    \rho(2)&=\frac{\gamma(2)}{\gamma(0)},~~ \mbox{where}\\
\gamma(2)&=\mbox{Cov}(Y_t,Y_{t-2})=E(Y_tY_{t-2})-E(Y_t)^2\\
         &=\sum_{i\in\mathcal{S}}\sum_{j\in\mathcal{S}} ij P(Y_t=j,Y_{t-2}=i)-E(Y_t)^2\\
         &=\sum_{i\in\mathcal{S}}\sum_{j\in\mathcal{S}} ij \sum_{h\in\mathcal{S}}P(Y_t=j|Y_{t-1}=h)P(Y_{t-1}=h|Y_{t-2}=i)P(Y_{t-2}=i)\\
         &-E(Y_t)^2\\
          &=\sum_{i\in\mathcal{S}}\sum_{j\in\mathcal{S}} ijP_{ij}^{(2)}\pi_i-E(Y_t)^2
\end{align*}
with $\pi_j = P(Y_t=j)$ i.e. the stationary distribution
where conditional probabilities come from the transition matrix (due to time-homogeneous assumption) and marginal probabilities from the stationary distribution (due to stationarity assumption). 
\end{comment}

    \item While for the general case of lag $k>1$ we get that
\begin{align*}
    \rho(k)&=\frac{\gamma(k)}{\gamma(0)},~~ \mbox{where}\\
\gamma(k)&=\mbox{Cov}(Y_t,Y_{t-k})=E(Y_tY_{t-k})-E(Y_t)^2\\
         &=\sum_{i\in\mathcal{S}}\sum_{j\in\mathcal{S}} ij P(Y_t=j,Y_{t-k}=i)-E(Y_t)^2\\
         &=\sum_{i\in\mathcal{S}}\sum_{j\in\mathcal{S}} ij \sum_{h_1\in\mathcal{S}}\sum_{h_2\in\mathcal{S}}\cdots \sum_{h_{k-1}\in \mathcal{S}} P(Y_t=j|Y_{t-1}=h_1)\\
         &P(Y_{t-1}=h_1|Y_{t-2}=h_2)\cdots P(Y_{t-k+1}=h_{k-1}|Y_{t-k}=i)P(Y_{t-k}=i)\\
         &- E(Y_t)^2\\
         &=\sum_{i\in\mathcal{S}}\sum_{j\in\mathcal{S}} ijP_{ij}^{(k)}\pi_i-E(Y_t)^2
\end{align*}
where conditional probabilities come from the transition matrix (due to time-homogeneous assumption) and marginal probabilities from the stationary distribution (due to stationarity assumption). 

\end{itemize}

\subsection{Autoregressive model of order $2$}
Let $\rho(k)$ denotes the auto-correlation fucntion at lag $k$ for a Markov chain of order $2$, $Y_t$ with state space $\mathcal{S}$. We consider that $Y_t$ is stationary, with stationary distribution $\pi$ and time homogeneous. Then, it holds that $E(Y_t)=E(Y_{t-k})$, while the transition matrix $M$ do not depend on time. In case of second order Markov Chain, the transition matrix $M$ has elements $M_{(\ell i),(ij)}$, where 
$$M_{(\ell i),(i j)}=P(Y_t=j|Y_{t-1}=i,Y_{t-2}=\ell),$$
while the stationary distribution $\pi$ consists of 
$$\pi_{ij}=P(Y_{t-1}=i,Y_t=j).$$

\begin{itemize}
   \item lag 0
\begin{align*}
\rho(0)&=1\\
\gamma(0)&=E(Y_t^2)-E(Y_t)^2\\
 &=\sum_{j\in \mathcal{S}}y_t^2P(Y_t=j)-(\sum_{j\in\mathcal{S}}y_tP(Y_t=j))^{2}\\
\end{align*}

\item lag 1
\begin{align*}
    \rho(1)&=\frac{\gamma(1)}{\gamma(0)},~~ \mbox{where}\\
\gamma(1)&=\mbox{Cov}(Y_t,Y_{t-1})=E(Y_tY_{t-1})-E(Y_t)^2\\
&=\sum_{i \in \mathcal{S}}\sum_{j \in \mathcal{S}}ij P(Y_t=j,Y_{t-1}=i)-E(Y_t)^2\\
&=\sum_{i\in \mathcal{s}}\sum_{j \in \mathcal{S}}ij \sum_{\ell \in \mathcal{S}}P(Y_t=j|Y_{t-1}=i,Y_{t-2}=\ell)P(Y_{t-1}=i,Y_{t-2}=\ell)-E(Y_t)^2\\
&=\sum_{i \in \mathcal{S}}\sum_{j \in \mathcal{S}}ij\sum_{\ell \in \mathcal{S}}M_{(\ell i)(i j)}\pi_{\ell j}-E(Y_t)^2
\end{align*}

\item lag 2
\begin{align*}
      \rho(2)&=\frac{\gamma(2)}{\gamma(0)},~~ \mbox{where}\\
\gamma(2)&=\mbox{Cov}(Y_t,Y_{t-2})=E(Y_tY_{t-2})-E(Y_t)^2\\
&=\sum_{i \in \mathcal{S}}\sum_{j \in \mathcal{S}}ijP(Y_t=j,Y_{t-2}=i)-E(Y_t)^2\\
&=\sum_{i \in \mathcal{S}}\sum_{j \in \mathcal{S}}ij \sum_{\ell \in \mathcal{S}}P(Y_t=j|Y_{t-1}=\ell,Y_{t-2}=i)P(Y_{t-1}=\ell,Y_{t-2}=i)-E(Y_t)^2\\
&=\sum_{i \in \mathcal{S}}\sum_{j \in \mathcal{S}}ij \sum_{\ell \in \mathcal{S}}M_{(i \ell )(\ell j)}\pi_{i\ell}-E(Y_t)^2
\end{align*}

\item lag 3
\begin{align*}
      \rho(3)&=\frac{\gamma(3)}{\gamma(0)},~~ \mbox{where}\\
\gamma(3)&=\mbox{Cov}(Y_t,Y_{t-3})=E(Y_tY_{t-3})-E(Y_t)^2\\
&=\sum_{i \in \mathcal{S}}\sum_{j \in \mathcal{S}}ijP(Y_t=j,Y_{t-3}=i)-E(Y_t)^2\\
&=\sum_{i \in \mathcal{S}}\sum_{j \in \mathcal{S}}ij P(Y_t=j,Y_{t-3}=i)-E(Y_t)^2\\
&=\sum_{i \in \mathcal{S}}\sum_{j \in \mathcal{S}}ij\sum_{\ell_1 \in \mathcal{S}}\sum_{\ell_2 \in \mathcal{S}}P(Y_t=j|Y_{t-1}=\ell_1,Y_{t-2}=\ell_2,Y_{t-3}=i)P(Y_{t-1}=\ell_1,Y_{t-2}=\ell_2,Y_{t-3}=i)-E(Y_t)^2\\
&=\sum_{i \in \mathcal{S}}\sum_{j \in \mathcal{S}}ij\sum_{\ell_1 \in \mathcal{S}}\sum_{\ell_2 \in \mathcal{S}}P(Y_t=j|Y_{t-1}=\ell_1,Y_{t-2}=\ell_2,Y_{t-3}=i)P(Y_{t-1}=\ell_1|Y_{t-2}=\ell_2,Y_{t-3}=i) \times \\
&P(Y_{t-2}=\ell_2,Y_{t-3}=i)-E(Y_t)^2\\
%&=\sum_{i \in \mathcal{S}}\sum_{j \in \mathcal{S}}ij\sum_{\ell_1 \in \mathcal{S}}\sum_{\ell_2 \in \mathcal{S}}M^{(2)}_{()()}\pi_{j\ell_2}-E(Y_t)^2
\end{align*}

\end{itemize}

%\textcolor{red}{last line need to check}

\end{document}